\documentclass[11pt,english]{article}
\usepackage{jheppub}

\usepackage{graphicx, float, array, xspace, amscd, amsmath, amsthm, amssymb, latexsym, bbm}
\newcommand{\be}{\begin{equation}}
\newcommand{\ee}{\end{equation}}
\newcommand{\bea}{\begin{eqnarray}}
\newcommand{\eea}{\end{eqnarray}}
\newcommand{\nn}{\nonumber}
\newcommand{\half}{\frac{1}{2}}
\newtheorem*{conjecture}{LVS Bootland Conjecture}
\newtheorem*{conjecture2}{AdS Bootland Conjecture}

\newcommand{\mbb}{\mathbb}
\newcommand{\mc}{\mathcal}
\newcommand{\ti}{\times}

\numberwithin{equation}{section}

\title{Putting the Boot into the Swampland}
\author[a]{Joseph P. Conlon,}
\author[b]{Fernando Quevedo}
\affiliation[a]{Rudolf Peierls Centre for Theoretical Physics, Clarendon Laboratory, University of Oxford, Parks Road, OX1 3PU, UK\\ 1 Keble Road, Oxford, OX1 3NP, England}
\affiliation[b]{ICTP, Strada Costiera 11, 34151 Trieste, Italy\\
DAMTP, CMS, University of Cambridge, Wilberforce Road, Cambridge, CB3 0WA, UK.Cambridge}
\emailAdd{joseph.conlon@physics.ox.ac.uk}
\emailAdd{f.quevedo@damtp.cam.ac.uk}
\abstract{The swampland program of delineating the space of effective field theories consistent with quantum gravity appears similar to the bootstrap program of delineating the space of quantum field theories consistent with conformal symmetry.
With this in mind we rewrite the effective field theory of the Large Volume Scenario in AdS space solely in terms of $R_{AdS}$, in a form suitable for holographic analysis. This rewritten EFT takes a remarkably universal (and previously unnoticed) form, which is uniquely determined in the large-volume limit up to terms suppressed by $\mc{O} \left( 1/\ln R_{AdS} \right)$,
with no reference to any of the fluxes, brane or instanton configurations that enter the microphysics of moduli stabilisation.
The putative dual 3d CFT will have two low-lying single trace scalars, an even-parity scalar $\Phi$ dual to the volume modulus with $\Delta_{\Phi} = \frac{3}{2}\left( 1 + \sqrt{19} \right) \simeq 8.038$ and an odd-parity scalar $a$ dual to the volume axion with $\Delta_a = 3$. On the AdS side the higher-point interactions are likewise uniquely determined. As the AdS theory is both subject to swampland constraints and holographically related to a CFT, we argue that holography will lead to a `bootland' --- a map between swampland constraints on the AdS side and bootstrap constraints on the CFT side. We motivate this with a discussion of swampland quantum gravity constraints on the axion decay constant in the $\mc{V} \to \infty$ limit and the $\langle \Phi \Phi a a \rangle$ 4-point function on the CFT side.
}

\begin{document}
\maketitle
\section{Introduction}

One of the oldest and most important tasks in string theory is to construct solutions that are as close as possible to the real
world -- solutions with four large dimensions, all moduli stabilised, and a matter spectrum containing
the Standard Model with the correct couplings.

It remains one of the most important tasks today. One central element to this program -- the question of moduli stabilisation -- has seen significant progress over the last fifteen years, with the construction of scenarios that (see \cite{DouglasReview, RalphReview, MaharanaPalti} for reviews) stabilise all moduli while also generating a hierarchical separation of scales that can result in interesting phenomenology. Nonetheless, these ideas have not been without criticism, and there remains a central question --- how can we \emph{know} that these scenarios are \emph{true} solutions of string theory?

The ordinary lines of argument for the existence and stability of these solutions use traditional physics methods. They are based on the techniques of dimensional reduction, effective field theory (EFT), expansions in small couplings, and explicit worldsheet computations of the effective action for the related supersymmetric Minkowski solutions. To maximise the degree of control, they often use 4-dimensional $\mc{N}=1$ supergravity as an effective theory to describe the compactification of 10d string theories, and then expand the K\"ahler and superpotentials in terms of the most important $\alpha'$ and $g_s$ corrections (see \cite{180808967, 180808971} for recent defenses of such tools). It is important to continue to develop such arguments as far as possible, making all details of the compactification and the effective theory explicit.

Nonetheless, these techniques only go so far. The main difficulty is that, in practice, the
constructions involve many moving parts, each of which may be complex. The techniques do not provide any
calculational equivalent of an \emph{experimentum crucis} that can definitively
verify the correctness of these scenarios. For example, in
Calabi-Yau compactifications preserving $\mc{N}=1$ supersymmetry at the compactification scale,
worldsheet calculations are intractable even in the absence of the fluxes or brane instanton effects required in semi-realistic scenarios.
Models \emph{without} supersymmetry suffer even more from the absence of a precise calculational framework.
This results in a proliferation of literature where an element of taste is required as to
how justified certain approximations are (for example, the large-N literature on the effects of anti-D3 branes down
warped Klebanov-Strassler throats in compact Calabi-Yaus).

This situation is not entirely satisfactory, when it is also claimed that many (or even all) of these solutions do not actually exist and instead belong to the swampland (for example \cite{hepth0412129, hepth0509212, 0601001, OoguriVafaI, OoguriVafaII, Sethi, Danielsson}).
While some aspects of the swampland critiques pertain only to de Sitter constructions, many parts are also relevant to AdS.
Furthermore, many EFT dS solutions are founded on
AdS solutions derived using a similar EFT logic. It is therefore simpler to ask first --- can we
determine whether the AdS solutions genuinely exist?

The lack of a precise calculational framework makes
the contours of the string swampland not well delineated. However, the contours of another swampland --- that
of 3d Conformal Field Theories (CFTs) --- have (in some cases) been much better mapped. Recent years have seen a resurgence of interest in the conformal bootstrap, together with powerful results constraining the allowed operator dimensions within such CFTs \cite{08070004, 12036064} (reviews are \cite{160207982, 180504405}).

In a strong form of the AdS/CFT correspondence, the 4d AdS solutions found in EFT string compactification will have CFT duals, implying the existence of 3d CFTs governed by specific expressions for the dimensions and couplings of low-lying operators.
The question of CFT duals of moduli-stabilised string vacua has attracted some attention
over the years (for earlier work see \cite{hepth0308175, hepth0412129, 08013326, 14126999}),
but it has appeared difficult to make quantitative predictions beyond very generic statements about
the existence of CFTs with low-lying scalars having $\Delta \sim \mc{O}(1)$ conformal dimension with a
gaps to higher (single-trace) modes.
Nonetheless, this makes the present a particularly apt time to reconsider these questions, and ask -- what would be the properties of such CFTs? Are they in the swampland? How do the swampland conjectures relate to CFT consistency conditions such as unitarity and crossing symmetry?

These questions define the philosophy of this paper --- to investigate the implications of $AdS_4$ moduli stabilisation scenarios and the swampland for putative 3d CFT duals. This paper will be particularly concerned with the Large Volume Scenario \cite{LVS, CQS}.
One reason for this is the existence of an attractive
large-volume $\mc{V} \ggg 1$ limit that is useful for analytic control
and also leads to many attractive phenomenological features.
However, we will also find that, when re-written in terms of $R_{AdS}$ instead of flux superpotentials, the
low-energy Lagrangian of the Large Volume Scenario exhibits a particularly simple (and previously unnoticed) form, leading to strikingly universal results for the conformal dimensions of dual operators and their interactions.

The paper is structured as follows. Section \ref{LVS2pt} reviews the Large Volume Scenario and discusses the spectrum of particle masses
as well as implications for the dimensions of dual operators in the CFT, splitting the analysis in terms of heavy and light modes.
 Section \ref{secHigherPoint} performs the same treatment for 3- and higher-point functions.
 Section \ref{others} makes a brief comparison to other models of AdS stabilisation within the framework of type IIB flux compactifications.
 Section \ref{BootlandSec} analyses the consequences of various swampland conjectures for the low-energy
 LVS Lagrangian and how these conjectures may manifest themselves within a dual 3d CFT language, although attempts at explicit
 CFT constructions of a dual to LVS are beyond the scope of this paper.

\section{Conformal Dimensions and 2-point functions in LVS}
\label{LVS2pt}

We start by reviewing the basic properties of the Large Volume Scenario (LVS) \cite{LVS, CQS}, which arises in the context of type D3/D7 IIB orientifolds with fluxes.
It is characterised by an AdS vacuum at exponentially large volume with broken supersymmetry. The supersymmetry breaking is approximately no-scale
and is inherited from the underlying
GKP model \cite{GKP} for flux stabilisation of complex structure moduli (in contrast to KKLT where the supersymmetric
K\"ahler moduli stabilisation eliminates the no-scale structure \cite{kklt}). The K\"ahler potential $K$ and superpotential $W$ \cite{GVW, DRS} are
\bea
K & = & - 2 \ln \left( \mc{V} + \xi \left( \frac{S + \bar{S}}{2} \right)^{3/2} \right) - \ln \left( -i \int \Omega \wedge \bar{\Omega} \right) - \ln \left( S + \bar{S} \right), \nn \\
W & = & \int G_3 \wedge \Omega + \sum_i A_i e^{-a_i T_i}.
\eea
Here $\mc{V}$ is the volume of the Calabi-Yau in units of $(2 \pi \sqrt{\alpha'})^6$, i.e. $\mc{V} = {\rm Vol} / (2 \pi \sqrt{\alpha'})^6$.
The superpotential is the sum of the flux superpotential $\int G_3 \wedge \Omega$ \cite{GVW, DRS}, with $G_3 = F_3 - S H_3$ a combination of the dilaton modulus ($S = \frac{1}{g_s} + ia_0$) and the RR and NS-NS 3-form fluxes. 
This depends implicitly on the complex structure ($U$) moduli via the holomorphic (3,0) form $\Omega (U)$. Finally, the superpotential also has a sum over effects non-perturbative in the K\"ahler moduli ($T = \tau + ia$) \cite{kklt} (either brane instantons or gaugino condensation). The K\"ahler potential arises from dimensional reduction of 10-dimensional type IIB supergravity, including the effect of the $\alpha'^{3}$ $\mc{R}^4$ correction \cite{CandelasDeLaOssa, BBHL}. Here $\xi = \frac{\zeta(3) \chi(M)}{2 (2 \pi)^3}$ is the coefficient of the $\alpha'^{3}$ correction (the relative factor of $g_s^{-3/2}$ in the $\left( \alpha' \right)^3$ term is because the volume is measured in Einstein frame and so already contains a $g_s^{-3/2}$ factor compared to string frame).%%JC

The Large Volume Scenario requires at least two K\"ahler moduli --- a `large' modulus controlling the overall volume and a `small' modulus corresponding to a blow-up cycle. The simplest form for the volume is
\be
\mc{V} = \frac{1}{\kappa} \left( \tau_b^{3/2} - \tau_s^{3/2} \right),
\ee
where $\kappa$ is a numerical constant that depends on the Calabi-Yau ($\kappa = \frac{1}{9\sqrt{2}}$ for the canonical LVS example of $\mbb{P}^4_{[1,1,1,6,9]}$ \cite{LVS}), $\tau_b = \half (T_b + \bar{T}_b)$ is the large cycle, and $\tau_s =
\half \left( T_s + \bar{T}_s \right)$ is the small cycle. Such a form for the volume is referred to as a `Swiss cheese' Calabi-Yau, with the small cycle viewed as making a hole in the cheese.

The dilaton and complex structure moduli are stabilised by fluxes by solving $D_{U, S} W =0 $.
Integrating these out, the simplest realisation of LVS has an effective theory for the K\"ahler moduli of
\bea
K & = & - 2 \ln \left( \frac{1}{\kappa} \left( \left( \frac{T_b + \bar{T_b}}{2} \right)^{3/2} - \left( \frac{T_s + \bar{T_s}}{2} \right)^{3/2} \right)  + \frac{\xi}{g_s^{3/2}} \right), \nn \\
W & = & W_0 + A_s e^{-a_s T_s}.
\eea
Computing the $\mc{N}=1$ supergravity scalar potential and extremising w.r.t $a_s = {\rm Im} (T_s)$, one obtains the standard LVS potential
\be
\label{LVSpotential}
V = \frac{A a_s^2 \sqrt{\tau_s} e^{-2 a_s \tau_s}}{\mc{V}} - \frac{B W_0 a_s \tau_s e^{-a_s \tau_s}}{\mc{V}^2} + \frac{C \xi W_0^2}{g_s^{3/2} \mc{V}^3},
\ee
where $A$, $B$, and $C$ are constants whose numerical values are unimportant here.

This potential has a minimum at exponentially large values of the volume with
\bea
\langle \tau_s \rangle & \sim & \frac{\xi^{2/3}}{g_s}, \nn \\
\langle \mc{V} \rangle & \sim & e^{a_s \langle \tau_s \rangle}.
\eea
The stabilised volume is exponentially dependent on $\frac{1}{g_s}$, allowing small variations in $g_s$ to lead to large differences in $\langle \mc{V} \rangle$. The exponentially large values of the stabilised volume allows us to view LVS as an approximation about a $\mc{V} \to \infty$ limit, with corrections parametrised by powers of $\mc{V}^{-1}$.

As $T_b$ does not enter the superpotential, the vacuum manifestly breaks supersymmetry as $F_{T_b} \neq 0$.
LVS preserves to leading order the no-scale structure of GKP models, and so in terms of symmetry breaking it inherits many of the properties of no-scale models (including soft terms and moduli masses that are much lighter than the `natural' value of the gravitino mass $m_{3/2}$ \cite{BCKMQ} --- see \cite{Reece} for a field theory explanation of this).%%JC

A distinguishing feature of LVS, crucial from a holographic perspective, is the fact that the only moduli
parametrically lighter than the gravitino mass are the volume modulus and its axion partner (the axion is a
pseudoscalar and the volume modulus is the pseudo-Goldstone boson of the broken scaling symmetry of no-scale supergravity).
All other moduli have masses that are comparable to, or larger than, the gravitino mass. As we will see, this implies that such moduli have large conformal dimensions when considered in the context of AdS/CFT.\footnote{Note the important contrast with `generic' supergravity models with many moduli, where $V \sim m_{3/2}^2 M_P^2$, $R_{AdS} \sim \frac{1}{m_{3/2}}$ and $m_{modulus} \sim m_{3/2}$, leading to many operators with $\mc{O}(1)$ conformal dimension.}

\subsection{Light Modes in LVS}

Following this telegraphic review of LVS, we now
discuss the light modes in LVS. `Light' here refers to any mode where the dual operator would have a conformal dimension that remains finite in the limit $\mc{V} \to \infty$ of LVS (some of the results here previously appeared in \cite{14126999}).

\subsubsection{Graviton}

We start with the `trivial' universal light mode on the AdS side. This is the 4-dimensional massless graviton $g_{\mu \nu}$, dual to the CFT stress tensor $T_{\mu \nu}$ with conformal dimension $\Delta = 3$.

\subsubsection{Volume and Axion Moduli}

We require an effective theory for these two light moduli. For the volume axion, the potential vanishes (although as we see
in Section \ref{secHigherPoint} the kinetic terms involve non-trivial interactions).
For the LVS volume modulus, the effective potential for the canonically normalised field takes the form
\be
\label{LVSeffPotential}
V = V_0 e^{- \lambda \Phi / M_P} \left( -\left( \frac{\Phi}{M_P} \right)^{3/2} + A \right),
\ee
(although we subsequently put $M_P$ =1). Here $V_0$ and $A$ are particular constants that depend on the microphysics of the compactification, while $\lambda = \sqrt{\frac{27}{2}}$ in LVS.
This potential arises as follows:
\begin{enumerate}
\item
Starting with the standard LVS potential of Eq. (\ref{LVSpotential}), we
solve for $\frac{\partial V}{\partial \tau_s} = 0$ and eliminate the heavy mode $\tau_s$.
This gives an effective potential for the volume modulus,
\be
\label{effPotential}
V_{eff} = \frac{1}{\mc{V}^3} \left( - A^{'} \left( \ln \mc{V} \right)^{3/2} + \frac{C'}{g_s^{3/2}} \right).
\ee
(this has used $e^{-a_s \tau_s} \sim \frac{\sqrt{\tau_s}}{\mc{V}}$ at the minimum).
This potential clearly has a minimum at exponentially large volume, the precise location of which depends on $A^{'}$ and $C^{'}$.
\item
The standard K\"ahler potential for the volume modulus is
\be
K = - 3 \ln (T_b + \bar{T}_b),
\ee
which results in a kinetic term
\be
\frac{3}{4 \tau^2} \partial_{\mu} \tau \partial^{\mu} \tau,
\ee
where $\tau = {\rm Re} (T_b)$.
\item
As a consequence the canonically normalised field is
\be
\Phi = \sqrt{\frac{3}{2}} \ln \tau = \sqrt{\frac{2}{3}} \ln \mc{V} + c
\ee
using the fact that $\mc{V} \propto \tau^{3/2}$.
\end{enumerate}
Writing (\ref{effPotential}) in terms of
the canonically normalised field $\Phi$ then gives the claimed potential. While this argument has neglected any mixing between $\tau_s$ and $\tau_b$ modes in integrating out $\tau_s$, such mixing is volume-suppressed and so can self-consistently be neglected.

We now analyse the implications of the potential of Eq. (\ref{LVSeffPotential}) from a holographic perspective.
For maximum generality, we do not specify $\lambda = \sqrt{\frac{27}{2}}$ yet, and
also phrase our analysis in terms of the more general potential
\be
V = V_0 e^{-\lambda \Phi} \left( - \Phi^{n} + A \right),
\label{hij}
\ee
which for appropriate values of $A$ has a minimum at large $\Phi \gg 1$ (equivalently, $\Phi \gg M_P$). It follows trivially that
\bea
V^{'} & = & V_0 e^{-\lambda \Phi} \left( -\lambda \left( - \Phi^n + A \right) - n \Phi^{n-1}  \right), \nn \\
V^{''} & = & - \lambda V^{'} + V_0 e^{-\lambda \Phi} \left( \lambda n \Phi^{n-1} \left( 1 - \frac{n-1}{\lambda} \frac{1}{\Phi} \right) \right).
\eea
At the minimum of the potential where $V^{'} = 0$, $\langle \Phi \rangle^{n-1} = -\frac{\lambda}{n} ( - \langle \Phi \rangle^n + A)$. Therefore
\bea
V^{''}_{min} & = & -\lambda^2 V_0 e^{-\lambda \Phi} \left( \left(-\langle \Phi \rangle^n + A \right) \left(1 - \frac{n-1}{\lambda} \frac{1}{\langle \Phi \rangle} \right) \right) \nonumber \\
& = & - \lambda^2 V_{min} \left( 1 - \frac{n-1}{\lambda} \frac{1}{\langle \Phi \rangle} \right),
\label{minprops}
\eea
where $\langle \Phi \rangle = \Phi \vert_{min}$.
As $\Phi \sim \ln \mc{V}$, the subleading correction is suppressed in the large volume limit.
We note two equivalent but illuminating
ways to rewrite Eq. (\ref{minprops}),
\bea
V^{''}_{min} & = & = 3 \frac{\lambda^2}{R_{AdS}^2} \left( 1 + \mc{O}\left(\frac{1}{\ln \left( R_{AdS}/l_P \right) } \right) \right), \\
V^{''}_{min} & = & 3 \frac{\lambda^2}{R_{AdS}^2}\left( 1 + \mc{O}\left(\langle g_s \rangle \right) \right).
\eea
Here $R_{AdS}$ represents the AdS radius of the 4-dimensional space-time. For a fixed (negative) 
cosmological constant $\Lambda$ this is given by $R_{AdS}^2 = - \frac{3 M_P^2}{\Lambda}$. For a dynamical minimum of the potential
$V_{min}$ we have therefore used %%JC
$V_{min} = -3 M_P^2 R_{AdS}^{-2}$ and $\langle \mc{V} \rangle \sim e^{\xi/g_s}$.

We now consider the implications for holographic conformal dimensions.
The general $AdS_4/CFT_3$ relationship between the conformal dimension of a dual scalar operator and the mass of the AdS excitation is
\be
\Delta (\Delta - 3) = m^2 R_{AdS}^2.
\ee
A conformal dimension $\Delta$ for an operator $\mc{O}$ is equivalent to a 2-point function $\langle \mc{O}(x) \mc{O}(y) \rangle = \frac{1}{\vert x - y \vert^{2 \Delta}}$. For the operator $\mc{O}_{\Phi}$ dual to the volume modulus, we then have
$
\Delta_{\Phi} (\Delta_{\Phi} - 3) = 3 \lambda^2 \left( 1 + \mc{O}\left( \frac{1}{\ln \mc{V}} \right) \right),
$
and so
\be
\Delta_{\Phi} = \frac{3\left(1 \pm \sqrt{1 + \frac{4}{3} \lambda^2}\, \right)}{2}\left( 1 + \mc{O}\left( \frac{1}{\ln \mc{V}} \right) \right).
\label{ConfDimension}
\ee
An attractive aspect of Eq. (\ref{ConfDimension}) is that, in the limit of asymptotically
large volumes, the scaling dimension of the dual operator to the light volume modulus $\Phi$ is uniquely determined.

Specialising to the case of LVS vacua with $\lambda = \sqrt{27/2}$ and $n=3/2$, Eq. (\ref{ConfDimension}) gives
\be
\label{LVSConfDimension}
\Delta = \frac{3 (1 + \sqrt{19})}{2} \left( 1 - \sqrt{\frac{2}{27}} \frac{1}{\langle \Phi \rangle} + \mc{O}\left( \frac{1}{\langle \Phi \rangle }\right)^2 \right).
\ee
Although we have arrived at Eq. (\ref{LVSConfDimension}) by analytically integrating out the small $\tau_s$ modulus, we have also
directly validated it by considering the full 2-modulus potential and numerically diagonalising and solving for the mass of
the light modulus.

As the axionic partner $a$ of the volume modulus has a mass suppressed by effects non-perturbative in
volume ($V(a) \propto e^{- b \mc{V}^{2/3}}$, for some constant $b$), to all practical purposes the mass vanishes. As this implies $\Delta_a (\Delta_a -3) = 0$, the dual operator to such an axion is exactly marginal with
\be
\Delta_a = 3.
\ee
The volume modulus is a scalar, while its axion partner is a pseudoscalar. In a dual CFT the volume modulus therefore corresponds to an even-parity scalar operator, whereas the dual of the axion field will be an odd-parity scalar operator. The graviton corresponds to an even parity spin 2 field.

If LVS vacua exist, they then correspond to a series of conformal field theories in which the dimension of the low-lying scalar operator dual to the volume has a conformal dimension (the inequality comes from Eq. (\ref{LVSConfDimension}))
\be
\Delta \leq 8.038 = \frac{3 (1 + \sqrt{19})}{2},
\ee
with the series of CFTs terminating at this asymptotic value for the conformal dimension. The central charge of these CFTs will behave as $c \sim \mc{V}^{3}$, being set by $V \vert_{min}$.

The asymptotic value is obtained in the limit of
infinitely large volume, in which the mass gap to other modes becomes arbitrarily large. Although in principle the size of the volume can only take discrete values, as $\ln \mc{V} \propto \frac{1}{g_s}$, and $g_s$ is fixed by the (very large) choice of fluxes, in practice it is reasonable to view the allowed values for $\mc{V}$ as a continuum attaining arbitrarily large values. For example, with $10^{200}$ flux choices satisfying the tadpole constraint, $g_s$ can be reasonably tuned to be as small as $10^{-100}$, allowing compactification volumes as
large as $e^{10^{100}}$.\footnote{While volumes larger than $10^{30}$ (in units of $l_s^6$) would be grossly inconsistent
with phenomenology as they imply $M_s < 1 {\rm TeV}$, such constraints are irrelevant for the questions of principle relevant to this paper.}

We can therefore regard LVS vacua as providing a series of vacua that approach the flat-space limit of AdS. In this limit,
the dimensions of low-lying primary single-trace operators in the CFT are shown in table \ref{TableOpDims}.
Note that, while the lowest \emph{single-trace} positive parity scalar operator has conformal dimension 8.038, there are \emph{double-trace} scalar modes with smaller conformal dimensions - both $T_{\mu \nu} T^{\mu \nu}$ and $a^2$ will have conformal dimension 6.
\begin{table}[]
\centering
\def\arraystretch{1.2}
\begin{tabular}{c | c | c | c}
\hline
Mode & Spin & Parity & Conformal dimension \\ \hline\hline
$T_{\mu \nu}$   & 2   & + & 3   \\
$a$   & 0             & - & 3    \\
$\Phi$ & 0 & + & 8.038  =  $\frac{3}{2} \left( 1 + \sqrt{19} \right)$\\
\hline
\end{tabular}
\caption{The low-lying single-trace operator dimensions for CFT duals of the Large Volume Scenario in the limit $\mc{V} \to \infty$.}
\label{TableOpDims}
\end{table}

\subsubsection{Fibre Moduli}

One interesting variation of the Large Volume Scenario is the case where there are many `large' moduli.
A simple example of this would be where the bulk space is a toroidal product, $T^2 \ti T^2 \ti T^2$.
More generally, this occurs for scenarios of fibred Calabi-Yaus \cite{08080691, 11070383}, for example when the volume can be expressed as
\be
{\mc V} = \alpha \left( \tau_1 \sqrt{\tau_2} - \tau_3^{3/2} \right).
\ee
In these models the overall volume direction is stabilised as in the normal Large Volume Scenario, by an interplay of $\alpha'$ and non-perturbative effects. This leaves the fibre direction (a simultaneous variation in $\tau_1$ and $\tau_2$ that leaves the overall volume unchanged) unstabilised. The fibre direction can be stabilised by perturbative D-brane loop corrections that break the additional degeneracy.
Such loop corrections \cite{BHKI,0507131,BHKII,CCQ} generate potential terms that are parametrically smaller than
the $\mc{O}(\alpha'^3)$ effects responsible for volume stabilisation, scaling instead as
$$
V_{loop} \propto \frac{1}{{\mc V}^{10/3}},
$$
smaller by a factor of $\mc{V}^{1/3}$ than the $V_{\alpha'^3}$ term. These generate a mass for the
fibre modulus that is lighter than the volume modulus by a factor of $\mc{V}^{1/6}$,
\be
M_{fibre} \propto \frac{1}{\mc{V}^{5/3}},
\label{fibremass}
\ee
while the corresponding axion field is again massless (as the stabilisation occurs via perturbative corrections to the K\"ahler potential which respect the axion shift symmetry). For a conventional analysis of moduli physics as manifested in cosmology or particle physics,
we must face the fact that the prefactor in Eq. (\ref{fibremass}) is highly model-dependent --- it is determined by matters such as the rank of the D7 brane gauge group and the precise cycle it wraps.
However, viewed from a holographic perspective this question simplifies.
Although the relative factor of $\mc{V}^{1/6}$ is not a large power, and the unknown
prefactors to the fibre modulus mass may be more important at small or moderate values of the volume, in the limit of asymptotically
large volume all such prefactors will be subdominant and the $\mc{V}^{1/6}$ factor will become parametrically large.
In this limit, it is therefore always true that
\be
\frac{M_{fibre}}{M_{volume}} \to 0,
\ee
and so, in particular,
\be
M_{fibre} R_{AdS} \to 0.
\ee
This implies that, within any holographic dual of a fibred version of LVS, the
conformal dimension of the operators dual to fibre moduli asymptote to $\Delta = 3$ for both the
fibre modulus and its axionic partner.

\subsubsection{Other Model-Dependent Light Modes}

The above has describe the minimal fields required for an LVS construction.
LVS is normally used as a starting point for phenomenological model building.
In any quasi-realistic LVS scenario, there will be additional light degrees of freedom, in particular a matter sector which
contains both chiral fermions and massless vector bosons and is used as a proxy for the Standard Model matter content. This sector may also contain light charged scalars, but the masses of these are more sensitive to the subtle details of supersymmetry breaking \cite{CQS, CAQS}, although it is expected that charged scalars will also satisfy $M^2_i \lesssim M_{\tau_b}^2$ \cite{BCKMQ, Reece}.

Such modes would be either light or massless within AdS space and so would correspond to additional spin-1/2 or spin-1 operators with conformal dimensions $\Delta \sim \mc{O}(1)$. Certainly, the presence of many such modes would make the analysis of the properties of any dual CFT far more intricate. However, from the perspective of moduli stabilisation these modes all appear to be optional.
LVS does not, by itself, require an interesting matter sector.
For this paper, we therefore assume that any dual CFT should not \emph{require} such modes for consistency and do not consider them further.
The universal aspect of LVS is associated to the dynamics of the volume modulus; if it is essential to consider the detail of D-brane model building within a compactification, any progress on CFT duals would seem intractable.

\subsection{Heavy Modes in LVS}

We now discuss the spectrum of heavy modes. Here `heavy' refers to modes for which the conformal dimension of the dual operator diverges
in the asymptotic limit of $\mc{V} \to \infty$.
These modes have masses larger than the volume modulus by some power of $\mc{V}$ (the exact power depending on the type of mode).
We restrict ourselves to states present within 10-dimensional supergravity and do not consider string states, although the discussion of KK states can be straightforwardly generalised to the case of string states.

\subsubsection{Small K\"ahler Modulus}

In addition to the volume modulus, a key role is played in LVS by the `small' cycle $\tau_s$. This field is crucial for volume
stabilisation because, as described in Eq. (\ref{LVSpotential}), it is non-perturbative effects in $T_s$ that lead to the $\left( \ln \mc{V} \right)^{3/2}$ term in the potential that balances against the $\alpha'^{3}$ correction to produce the minimum at exponentially large volumes.

However, in the vacuum the $\tau_s$ mode is parametrically heavier than the volume modulus. The
mass of this mode is \cite{07053460}
\be
m_{\tau_s} \simeq 2 m_{3/2} \ln \left( M_P / m_{3/2} \right) \sim \frac{M_P \ln \mc{V}}{\mc{V}}.
\ee
As the mass arises from non-perturbative effects, the axion and fermion partners also have similar masses and so
in terms of conformal dimension, all these modes correspond to operators with dimension
$$
\Delta_{\tau_s} \sim (\ln \mc{V}) {\mc{V}}^{1/2} \gg 1.
$$

\subsubsection{Dilaton and Complex Structure Moduli}

In IIB flux compactifications, the dilaton ($S$) and complex structure ($U$) moduli are stabilised
by the 3-form fluxes through a superpotential \cite{GVW, DRS}
\be
W = \int G_3 \wedge \Omega,
\ee
together with the K\"ahler potential \cite{CandelasDeLaOssa}
\be
K_{S,U} = - \ln \left( i \int \Omega (U) \wedge \bar{\Omega} (\bar{U}) \right) - \ln \left( S + \bar{S} \right).
\ee
Here $G_3$ is the complexified 3-form $G_3 = F_3 + S H_3$,
$\Omega$ the holomorphic $(3,0)$ form of the Calabi-Yau and $S = \frac{1}{g_s} + i a_0$ the IIB
dilaton-axion multiplet.

The precise number of such moduli depends on the topology of the Calabi-Yau and can vary from $\mc{O}(1)$ to $\mc{O}(1000)$.
The supergravity scalar potential is
\be
V = e^K \left( K^{i \bar{j}} D_i W D_{\bar{j}} \bar{W} - 3 \vert W \vert^2 \right).
\ee
Including $- 2 \ln (\mc{V})$ from the K\"ahler potential, the effective potential for these moduli is
\be
V = e^{K_{S,U}} \frac{K^{i \bar{j}} D_i W D_{\bar{j}} \bar{W}}{\mc{V}^2},
\label{cspotential}
\ee
where the indices $i,j$ run over the dilaton and complex structure moduli. Although no-scale is not exact, the terms that break no-scale and give rise to LVS are subleading in the K\"ahler potential and at $\mc{O}(\mc{V}^{-3})$, and so can be neglected when considering stabilisation of the dilaton and complex structure moduli (see \cite{CQS} for a more detailed discussion of this point).

The potential of Eq. (\ref{cspotential}) is minimised by solving the first-order equations $D_{U,S} W = 0$ with vanishing F-terms for $U$ and $S$ moduli. This produces a characteristic mass scale for the dilaton and complex structure moduli,
\be
m_{U,S} \sim \frac{M_P}{\mc{V}},
\ee
where we only stress the volume dependence of the mass.
As the stabilisation of these moduli is supersymmetric, the same mass scaling holds for the fermionic partners.

The scaling dimension of the operators within the dual $\mc{N}=1$ superconformal multiplets is then
$$
\Delta_{U, S} \sim \mc{V}^{1/2} \gg 1,
$$
and so parametrically decoupled in the $\mc{V} \to \infty$ limit.

\subsubsection{Gravitino and Modulini}

A further universal mode is the gravitino $\psi_{3/2}$
with a mass set by $m_{3/2} = e^{K/2} \vert W \vert$. Using $K(T + \bar{T}) = - 2 \ln \mc{V}$, this gives the simple result
\be
m_{3/2} = \frac{W_0 M_P}{\mc{V}},
\ee
and so the conformal dimension of the dual spin 3/2 operator is
$$
\Delta_{3/2} \sim \mc{V}^{1/2} \gg 1.
$$
Similar results hold for all modulini since once supersymmetry is broken their mass is of order of the gravitino mass (see for instance \cite{deCarlos:1993wie}).

\subsubsection{KK Modes}

We now consider heavy modes present in the 10-dimensional theory but not in the 4-dimensional one.
An important and universal
set of such modes are the KK modes from
dimensional reduction of the 10d supergravity theory down to four dimensions. They arise from solving (for scalars)
\be
g^{\mu \nu} \partial_{\mu} \partial_{\nu} \phi = m^2 \phi,
\ee
with all derivatives evaluated within the compact dimensions.

The lightest KK modes start at
\be
M_{KK} \sim \frac{M_s}{R} \sim \frac{M_P}{\mc{V}^{2/3}}.
\label{KKmass}
\ee
Eq. (\ref{KKmass})
has a prefactor that depends on the exact geometry (as in $l(l+1)$ for spherical harmonics) but our focus is simply the power of volume $\mc{V}$ that appears.

Starting with the lowest KK modes, a tower of excited KK harmonics is built up.
For instance, for a six-dimensional torus, the masses of the states in the tower are
\be
m_i^2 = \left( n_1^2 + n_2^2 + n_3^2 + n_4^2 + n_5^2 + n_6^2 \right) M_{KK}^2 \equiv R_{KK}^2 M_{KK}^2, \qquad \textrm{with} \, \, n_i \in \mbb{Z}.
\ee
Here $R_{KK} \equiv \sqrt{n_1^2 + n_2^2 + n_3^2 + n_4^2 + n_5^2 + n_6^2}$ denotes the radius in `KK number' space.
This corresponds to a spectrum of progressively heavier states with spectral density\footnote{We expect this relationship to hold
in the large $\mc{V}$ limit for all approximately homogeneous spaces.}
\be
\frac{dN}{dR_{KK}} \propto R_{KK}^5.
\ee
In general, an effective field theory including only the KK modes has a natural cutoff at the string scale, when string modes must also be included.
This cutoff occurs when $R_{KK} \sim {\mc{V}^{1/6}}$ and so there will be $\mc{O}(\mc{V})$ KK states in the tower below the string scale.

Considered within a dual CFT, this tower corresponds to a tower of high-dimension operators with
\be
\Delta_i \simeq R_{KK} \mc{V}^{5/6}
\ee
and whose number density for $1 \lesssim R_{KK} \lesssim \mc{V}^{1/6}$ behaves as
\be
\frac{dN}{d R_{KK}} \propto R_{KK}^5.
\label{specKK}
\ee
The spectral density of Eq. (\ref{specKK}) holds for $\Delta \lesssim \mc{V}$.
For conformal dimensions $\Delta \gtrsim \mc{V}$, the number density of operators grows exponentially as operators dual to excited string states appear.

As the KK spectrum comes from dimensional reduction of all modes in the 10-dimensional theory, it
contains not just scalars, but also particles with spins 1/2, 1, 3/2 and 2, coming from dimensional reduction of the graviton and gravitino.
As the scale of supersymmetry breaking is much smaller than the KK scale, we expect these modes (and the dual operators) to arrange into
complete supersymmetry multiplets.

\subsection{Stability of the AdS solution and Holography}

One general question sitting in the background here is whether metastable or non-supersymmetric AdS solutions can be meaningfully described from a holographic perspective (indeed, one of the conjectures of \cite{OoguriVafaII} (also see \cite{161004564}) is that non-supersymmetric AdS solutions do not exist (see however \cite{Giombi:2017mxl})). The argument goes as follows: if there exists a Coleman-de Luccia (CdL) bubble that can mediate a transition to a more stable endpoint then there exists a finite probability per unit volume for this decay to occur. If this holds, the combination of the infinite volume and
causality structure of AdS implies that decays percolate in from the boundary and so any observer feels the decay within an AdS time, no matter how suppressed the decay probability is \cite{07094262, 10035909}.

There are two main points to make about the relationship of this argument to the case at hand.
The first is that, at least within its 4-dimensional effective field theory, LVS is an absolute minimum of the scalar potential rather than a
metastable one --- there is no other AdS minimum anywhere within the field space. Furthermore, the potential decay of one AdS vacuum to another with different fluxes (and superpotential) in LVS was studied in \cite{deAlwis:2013gka} where it was found that the tension of a nucleating five-brane between two different AdS vacua is too large compared with the difference in their respective vacuum energies to allow for the nucleation to happen. Of course, these statements are defined only within the 4d effective field theory. The string landscape is believed to be continuously connected, and there are certainly stable AdS solutions of string theory with deeper potentials --- for example $AdS_4 \ti S^7$ solutions of M-theory. While these are very far from LVS, and would involve multiple changes in topology and flux configuration, as well as passing through a strong coupling transition to make the eleventh dimension geometric, the belief is that they would be connected within the fundamental quantum gravity theory.

If this is true, as the $\mc{V} \gg 1$ limit of LVS is strongly classical, the distance in field space from the LVS to another AdS vacua with $V_{min} < V_{LVS, min}$ would nonetheless be expected to be $\Delta \Phi \gg M_P$.
This reasoning makes it possible that some highly generalised version of a CdL instanton or bubble of nothing could exist, but one should not speculate here too much in the absence of any concrete computation.

The second point is that intrinsic to the argument against non-supersymmetric holography is the idea that there is no qualitative difference between a decay amplitude of $e^{-10^{100}}$ and a decay amplitude that is $\mc{O}(1)$, as each is multiplied by an infinite volume.
However such an argument must be treated with a lot of caution, as it is unusual in physics that quantities that are arbitrarily small cannot be usefully approximated to zero (for example, the proton decay rate). Also, despite infinite volumes, relevant quantities are usually decay rates per unit time and unit volume. It is also well established that non-conformal theories with running couplings (for example, QCD) can be sensibly regarded as approximately conformal, with the conformal symmetry group used to derive useful results.

Finally, the main argument in \cite{OoguriVafaII} was based on examples of non-supersymmetric flux-stabilised vacua, obtained as a near-horizon limit of brane configurations. This is a very different origin for non-supersymmetric AdS than the interplay of perturbative and non-perturbative corrections which are the basis of the LVS solutions. So while we note these arguments against non-supersymmetric holography, we will still proceed with a standard analysis of the consequences of the AdS LVS solution for its putative holographic dual $CFT_{LVS}$.

\section{3-pt and Higher Point Functions}
\label{secHigherPoint}

We now consider the higher-point functions that are present within AdS space. We structure our analysis by considering first higher-point couplings that only involve the light fields, and then extending to mixed couplings between heavy and light modes.
Within a dual CFT, these couplings will relate via Witten diagrams to the structure constants of the CFT \cite{Witten},
which play a crucial role for determining the consistency of the CFT with unitarity and crossing symmetry.

\subsection{Interactions Only Involving Light Modes}

We start with couplings involving the massless (equivalently $\Delta_a = 3$) pseudoscalar field $a$.
As it has no potential, the field $a$ has no $a^n$ self-couplings.

Nonetheless, it does  couple to the volume modulus $\Phi$ through its kinetic terms.
These arise from the kinetic terms associated to $K = - 3 \ln \left( T + \bar{T} \right)$,
\be
\mc{L} = \frac{3}{4 \tau^2} \partial_{\mu} \tau \partial^{\mu} \tau  + \frac{3}{4 \tau^2} \partial_{\mu} a \partial^{\mu} a.
\label{basekin}
\ee
After canonical normalisation of $\Phi = \sqrt{\frac{3}{2}} \ln \tau$, Eq. (\ref{basekin}) becomes
\be
\mc{L} = \frac{1}{2} \partial_{\mu} \Phi \partial^{\mu} \Phi + \frac{3}{4} e^{-\sqrt{\frac{8}{3}} \Phi} \partial_{\mu} a \partial^{\mu} a.
\label{akineticterm}
\ee
Eq. (\ref{akineticterm}) has a clear physical meaning --- in the limit $\langle \Phi \rangle \gg 1$, the field range $2 \pi f_a$
of the axion reduces, with
$$
f_a \sim \frac{M_P}{\mc{V}^{2/3}}
$$
in the large $\mc{V}$ limit (restoring factors of $M_P$).

Expanding about the minimum of the potential, $\Phi = \langle \Phi \rangle + \delta \Phi$, we obtain a canonically normalised axion field by redefining $a \to
\sqrt{\frac{3}{2}} e^{-\sqrt{\frac{2}{3}} \langle \Phi \rangle} a$. In terms of dynamical fluctuations about the vacuum and including factors of $M_P$, the interactions between $a$ and $\delta \Phi$ are then
\bea
\mc{L}_{3-pt} & = & - \sqrt{\frac{2}{3}} \left( \frac{\delta \Phi}{M_P} \right) \partial_{\mu} a \partial^{\mu} a, \nn \\
\mc{L}_{4-pt} & = & \frac{2}{3} \left( \frac{\delta \Phi}{M_P} \right)^2 \partial_{\mu} a \partial^{\mu} a, \nn \\
\textrm{and generally} \qquad \mc{L}_{n-pt} & = & \left( - \sqrt{\frac{8}{3}} \right)^{(n-2)} \frac{1}{2 (n-2)!}\left( \frac{\delta \Phi}{M_P}\right)^{n-2} \partial_{\mu} a \partial^{\mu} a.
\label{axinteractiona}
\eea
We note that the form of Eq. (\ref{axinteractiona}) is independent of any of the microphysics of the compactification (for example the flux choice, the value of $W_0$, the form of non-perturbative effects, etc).

Of course, Eq. (\ref{axinteractiona}) is only a leading approximation to the full LVS Lagrangian. In the full
Lagrangian there are additional contributions to these mixed couplings, arising (for example) from higher-derivative interactions always present in effective string Lagrangians. We can split these effects into two sorts.
First, effects on the internal space (such as the $\alpha'^{3}$ term used in LVS) alter the kinetic terms away from Eq. (\ref{basekin}),
modifying the canonical normalisation by additional volume-suppressed terms.
Alternatively, there are additional corrections directly in 4-dimensions, in particular those that extend the 4d action beyond the
two derivative level.

We can use a similar logic to neglect both sets of terms --- they each involve terms suppressed by higher powers of $R_{AdS}$.
Two derivative couplings in AdS space are of order the characteristic curvature scale,
 namely $R_{AdS}^{-2}$. As higher derivative terms involve higher powers of curvature, they scale as $R_{AdS}^{-4}$ or greater.
 Likewise, as $R_{AdS} \sim \mc{V}^{3/2}$, any extra-dimensional terms suppressed by additional powers of volume also give effects suppressed by powers of $R_{AdS}$. In the limit of large compactification volumes and large AdS radius, it is self-consistent to neglect all such terms compared to those of Eq. (\ref{axinteractiona}).

We now consider self-couplings of the volume modulus with itself.
Redefinition of the volume modulus to canonical form using $\Phi = \sqrt{\frac{3}{2}} \ln \tau$
eliminates any self-couplings involving the kinetic term (this is modulo volume-suppressed effects
associated either to higher order corrections to $K = - 2 \ln \mc{V}$ or to the mixing of $\tau_b$ and $\tau_s$ in the LVS potential --- but as such effects are subleading by $\mc{V}^{-1}$, it is again self-consistent to neglect them).

We are then left with
only the higher-point self-interactions in the scalar potential, in an expansion
\be
V = V_0 + \sum \frac{g_n}{n!} \left(  \frac{\delta \Phi}{M_P} \right)^n.
\ee
While we can determine $g_n$ recursively using similar arguments to that for Eq. (\ref{minprops}), there is a nicer way
to obtain a general expression for $g_n$, starting with the potential
\be
V = V_0 e^{-\lambda \Phi} \left( - \Phi^{k} + A \right).
\label{start}
\ee
The minimum of Eq. (\ref{start}) is at $\langle \Phi \rangle = \Phi_0$, where
\be
A-\Phi_0^k=-\frac{k}{\lambda} \Phi_0^{k-1},
\ee
at which
\be
V(\Phi_0)=-\frac{kV_0}{\lambda}\, e^{-\lambda \Phi_0} \Phi_0^{k-1}
\ee
Let us Taylor expand $V$ around $\Phi=\Phi_0$.
\bea
V&= & V_0e^{-\lambda\Phi_0} e^{-\lambda(\Phi-\Phi_0)}\left( A-\Phi_0^k\left(1+\frac{\Phi-\Phi_0}{\Phi_0}\right)^k\right)\nonumber \\
&\simeq &V_0e^{-\lambda\Phi_0} e^{-\lambda(\Phi-\Phi_0)}\left( A-\Phi_0^k-k\Phi_0^{k-1}(\Phi-\Phi_0)\right)\nonumber \\
&\simeq& -\frac{kV_0}{\lambda}\Phi_0^{k-1} e^{-\lambda \Phi_0} e^{-\lambda(\Phi-\Phi_0)}\left(1+\lambda(\Phi-\Phi_0)\right)\nonumber \\
&\simeq& V(\Phi_0)\left[\sum_n\frac{(-1)^n\lambda^n(\Phi-\Phi_0)^n}{n!} +\sum_m\frac{(-1)^m\lambda^{m+1} (\Phi-\Phi_0)^{m+1}}{m!}\right]\nonumber \\
& \simeq& V(\Phi_0)\sum_n(-1)^n\lambda^n(\Phi-\Phi_0)^n\left(\frac{1}{n!}-\frac{1}{(n-1)!}\right)\nonumber \\
&\simeq & V(\Phi_0)\sum_n\frac{(-1)^{n-1}(n-1)}{n!}\lambda^n(\Phi-\Phi_0)^n.
\eea
Since we can in general expand
\be
V(\Phi)=\sum_n\frac{V^{(n)}(\Phi_0) (\Phi-\Phi_0)^n}{n!},
\ee
we can read off
\be
V^{(n)}(\Phi_0)=(-1)^{n-1}(n-1)\lambda^n V(\Phi_0).
\ee

It therefore follows that (where $\lambda = \sqrt{\frac{27}{2}}$ for LVS) the n-point self-interaction of the volume modulus in AdS space is
\be
{\mc L}_{n-pt} = (-1)^{n-1} \lambda^n (n-1)  \left( - 3 \frac{M_P^2}{R_{AdS}^2} \right) \frac{1}{n!} \left( \frac{\delta \Phi}{M_P} \right)^n \left( 1 + \mc{O} \left( \frac{1}{\lambda \langle \Phi \rangle} \right) \right).
\label{volmodulus}
\ee

Let us make some comments about the forms of the self-interaction. First, the structure of Eq. (\ref{volmodulus}) is radiatively stable.
The fact that LVS arises from type IIB flux compactifications within an effective low-energy $\mc{N}=1$ supergravity theory implies
that quantum corrections must respect this microphysics. Quantum corrections have been studied extensively within LVS and, as a consequence of the extended no-scale structure, all known corrections give rise to terms in the potential that are
subleading by (fractional) powers of the volume \cite{BHKI,0507131,BHKII,CCQ}. As they do not modify the basic form of the LVS potential in Eq. (\ref{LVSpotential}), they leave the structure of Eq. (\ref{start}) unaffected, and so the form of  Eq. (\ref{volmodulus}) is
stable.\footnote{The reader may wonder as to the physics of why UV divergences from loops would not renormalise the couplings of Eq. (\ref{volmodulus}). The point is that, in string theory, $\langle \Phi \rangle$ itself sets the scale of the UV regulator, as $\langle \Phi \rangle$ determines the compactification volume, and so also both the string scale and the scale of supersymmetry restoration $m_{3/2}$. The $\Phi$ potential is therefore self-controlled, in the sense that the UV cutoff $\Lambda = \Lambda \left( \langle \Phi \rangle \right)$ itself ensures that UV divergences are regulated at a low enough scale to not affect the leading terms in $V(\Phi)$.}

Second, we see that the detailed microphysics does not enter Eq. (\ref{volmodulus}) except via the subleading $\frac{1}{\lambda \langle \Phi \rangle}$ terms. The leading part of Eq. (\ref{volmodulus}) is independent of $W_0$, instantons, $\chi(CY)$, $g_s$ or any of the other detailed properties of the compactification. This implies the attractive feature that in the asymptotic limit of $\mc{V} \to \infty$ all the interactions are uniquely determined and can be specified solely in terms of $R_{AdS}$.

At finite volumes, we see that both the conformal dimensions and higher-point interactions are corrected by an expansion in $\frac{1}{\lambda \langle \Phi \rangle}$. Similarly for the 2-point function, this expansion in $\frac{1}{\lambda \langle \Phi \rangle}$ is equivalent to an expansion in powers of $g_s$, which is appealing from a potential holographic interpretation. We note that the coefficients of these subleading corrections \emph{do} depend on the detailed microphysics of the geometry --- for example, $\langle \Phi \rangle$ depends on the value of $\xi$, which involves the Euler number of the Calabi-Yau.\footnote{We note that Calabi-Yau geometries have made an appearance in quantum field theory loop amplitudes \cite{180509326}.}

This suggests that the right way to think about the LVS solution, at least in a holographic sense, is as an expansion about the $\mc{V} \to \infty$ limit, organised in corrections of the form $\frac{1}{\ln \mc{V}}$. Universal behaviour occurs in the strict limit $\mc{V} \to \infty$; model-dependent corrections occurs at finite volume.

As with the axion kinetic terms, there will also be
further corrections to the interactions of Eq. (\ref{volmodulus}) that are suppressed by
actual powers of $\mc{V}$ (for example, higher $\alpha'$ corrections).
As corrections suppressed by powers of $\mc{V}$ are highly
subleading compared to those terms listed in Eq. (\ref{volmodulus}) that are suppressed by $\ln \mc{V}$, we do not consider them further.

Finally, we consider interactions involving the graviton mode, $g_{\mu \nu}$, that is dual to the stress tensor of the CFT.
These include its kinetic self-interactions, deriving from the Einstein-Hilbert term $\int \sqrt{g} \mc{R}$, and also the coupling of the graviton
to scalar field kinetic terms, via
$$
\int \half g_{\mu \nu} \partial^{\mu} a \partial^{\nu} a, \qquad \qquad \textrm{ or } \qquad \qquad
\int \half g_{\mu \nu} \partial^{\mu} \Phi \partial^{\nu} \Phi,
$$
where we have used $\Phi$ and $a$ to denote canonically normalised fields. They also include interactions with potential terms,
$$
\int d^4 x \sqrt{g} V(\Phi, a).
$$
As the structure of the graviton interactions is fixed by general relativity,
there is nothing particular to LVS in the form they take.

\subsection{Mixed interactions of Heavy Modes with Light Modes}

The volume axion does not appear in the potential, and its kinetic coupling only involves the light volume modulus. Up to higher order terms suppressed by powers of volume it therefore does not couple to any of the heavier moduli (such higher order terms could be induced, for example, by mixings between the K\"ahler and complex structure moduli --- this, however, being volume-suppressed compared to the tree-level K\"ahler potential).

The expectation value of the volume modulus itself, though, plays a major role in determining the masses of all heavier modes.
In a framework of Einstein-Hilbert gravity, the 4-dimensional Planck scale is a fixed constant of $M_P = 2.4 \ti 10^{18} \, {\rm GeV}$.
As $M_s \propto \frac{M_P}{\sqrt{\mc{V}}}$, variations in the expectation value of the volume modulus, $\langle \Phi \rangle$, physically correspond to variations in the string scale (as $M_s \propto \frac{M_P}{\sqrt{\mc{V}}}$) and so as $\langle \Phi \rangle$ varies, so does the mass of all heavy modes.

This feature determines the form of the couplings between the volume modulus and the heavy modes. We illustrate this by focusing on the examples of the flux-stabilised dilaton and complex structure moduli and also the KK modes.

Let us first discuss the $S$ and $U$ moduli. The kinetic terms for these fields have no dependence on the volume and so, in principle, canonical normalisation can be performed purely within the $(S, U)$ sector. From the potential of Eq. (\ref{cspotential}), we see that the volume enters as a universal prefactor of $\mc{V}^{-2}$.
This factor fixes the higher-point functions of the volume modulus with the $S$ and $U$ moduli, implying the
interaction with the volume modulus is of the form
\be
V \propto e^{-\sqrt{6} (\delta \Phi / M_P) } M^2_{i\bar{j}} U_i \bar{U}_{\bar{j}},
\label{def}
\ee
where $M^2_{i\bar{j}}$ is the physical mass squared matrix for complex structure moduli \emph{in the vacuum}. Expansion of Eq. (\ref{def}) then gives the couplings
\be
\mc{L}_{(\delta \Phi)^n U U^{*}} = \frac{1}{n!} \left( - \sqrt{6} \frac{\delta \Phi}{M_P} \right)^n M^2_{i\bar{j}} U_i \bar{U}_{\bar{j}}.
\ee

In principle, similar arguments can be used for the KK modes. These modes arise from eigenmodes of the Laplace operator in the extra-dimensional space,
\be
\partial_{\mu} \partial^{\mu} \phi = m^2 \phi.
\ee
The volume modulus corresponds to the isotropic rescaling mode $g_{\mu \nu} \to \lambda^2 g_{\mu \nu}$, and so it follows that the effect of this rescaling is to modify the mass of KK modes, $m^2 \to \lambda^{-2} m^2$. In particular, such rescalings have the same effect on \emph{all} of the KK modes - they keep the relative form of the KK modes unaltered, and act as a common rescaling on the mass and kinetic terms.\footnote{This holds even for models with $N \gg 1$ K\"ahler moduli --- the goldstino of the no-scale GKP model, inherited by LVS, is aligned perfectly with the isotropic rescaling mode \cite{07100873}.}

This allows the interaction of KK modes with the volume modulus to be determined, i.e. the interactions
$$
\sum a_n \left( \frac{\delta \Phi}{M_P} \right)^n \partial_{\mu} H \partial^{\mu} \bar{H} + m^2 \sum b_n \left( \frac{\delta \Phi}{M_P} \right)^n H \bar{H} \in \mc{L}.
$$
where $H$ denotes the KK mode.
To determine these, we consider what a KK mode would look like, when `integrated in' to a 4d supergravity description (see \cite{0611144, 13106694} for similar arguments). The shift symmetry of the K\"ahler moduli and the holomorphy of the superpotential implies that only the K\"ahler metric for the KK modes can explicitly depend on volume (while the KK modes must have a superpotential mass, the superpotential can have no explicit perturbative dependence on $T$ moduli).

We use $H$ to denote the (un-normalised) KK modes.
KK modes have $m_{KK}^2 \propto \frac{M_P^2}{\mc{V}^{4/3}}$. To ensure this behaviour,
and taking into account the $\mc{V}^{-2}$ from $e^K$, the kinetic terms for $H$ must scale as $\mc{V}^{-1/3} \equiv (\tau)^{-1/2}$. This implies a K\"ahler metric and superpotential for the KK modes
\bea
K & = & - 2 \ln \mc{V} + \frac{1}{(T + \bar{T})^{1/2}} H \bar{H} + \ldots, \nn \\
W & = & \ldots + M_{KK}(U) H^2 + \ldots
\eea
producing an interaction Lagrangian (including the mass terms)
\bea
\mc{L}_{KK} & = &
\frac{1}{(T + \bar{T})^{1/2}} \partial_{\mu} H \partial^{\mu} \bar{H} -
\frac{1}{2 (T + \bar{T})^{3/2}} \left( \partial_{\mu} T \partial^{\mu} \bar{H} H + \partial_{\mu} \bar{T} \partial^{\mu} H \bar{H} \right) \nn \\
& & + \frac{3}{4 (T + \bar{T})^{5/2}} \partial_{\mu} T \partial^{\mu} \bar{T} H \bar{H} + \frac{M_{KK}^2}{(T + \bar{T})^{5/2}} H \bar{H}.
\eea
By writing $H^{'} = \frac{H}{\tau^{1/4}}$ we can proceed to make the KK modes canonically normalised. In terms of $H'$ and $\Phi$ these couplings are of the form:

\be
M_{KK}^2e^{-\sqrt{8/3}\,\delta\Phi/M_P}\, H'\bar{H'},
\ee
and
\be
e^{-\sqrt{1/6}\, \delta\Phi/M_P}\left\{
\partial_{\mu} H' \partial^{\mu} \bar{H'},\quad \partial^\mu\left(\frac{\delta\Phi}{M_P}\right)\partial_\mu \left(\frac{\delta\bar{\Phi}}{M_P}\right) H'\bar{H'},\quad H' \partial^\mu\left(\frac{\delta\Phi}{M_P}\right)\partial_\mu \bar{H} +h.c\right\}
\ee
Expanding the exponentials as before we can read the direct couplings to $\left( \delta\Phi \right)^n$. A similar analysis can in principle be done for massive string modes.
%However, we do not pursue a detailed study of the interactions of the KK modes further here.

\section{Comments on non-LVS models}
\label{others}

In this section we make some brief comments about the implications of other non-LVS constructions for putative CFT duals.

\subsection{`LVS' with $\lambda \leq \sqrt{6}$}

The ordinary derivation of LVS gives $\lambda = \sqrt{\frac{27}{2}}$ as the unique correct value of $\lambda$ in Eq. (\ref{hij}).
However, it is also interesting to note the case $\lambda \leq \sqrt{6}$. The value of $\sqrt{6}$ is interesting because, as $m_s \propto \frac{M_P}{\sqrt{\mc{V}}}$, $\lambda = \sqrt{6}$ corresponds to a potential that scales at large volume as
\be
V \propto M_s^4 \sim \frac{M_P^4}{\mc{V}^2} \left( 1 + \ldots \right).
\ee
One may view a $V \propto M_s^4$ scaling as somewhat `natural', although we stress
that there is no known construction that both has this scaling and still gives a minimum of the
potential at asymptotically large volumes.

From this respect $\lambda < 6$ is also interesting for two reasons. First, as this corresponds in the perturbative weakly-coupled limit of
$\mc{V} \to \infty$ to $V \gg M_s^4$, one may suspect that it belongs to the swampland --- such a scaling would require brane configurations that are intrinsically non-supersymmetric and involve $D5/\bar{D}5$ pairs or $D7/\bar{D}7$ pairs (the energy of a $D3/\bar{D}3$ system only scales as $V \propto \mc{V}^{-2}$). Second, from a holographic perspective the case of $\lambda < 6$ also marks a clear transition, as it implies
$\Delta_{\Phi} < 6$ and so would make the volume operator the lowest-dimension even parity scalar operator, bringing it below $T_{\mu \nu} T^{\mu \nu}$ and $a a$.

\subsection{Perturbative Stabilisation}

We consider here models of perturbative stabilisation of the volume modulus (for example as in \cite{0507131, 0602253, BHKII}).
Typically these arise from balancing effects of different order in the volume expansion --- e.g. balancing the $\mc{V}^{-3} \alpha'^{3}$ correction
against $\mc{V}^{-10/3}$ string loop corrections. In terms of the canonically normalised field $\Phi$, for such models we can write, allowing arbitrary values for the two powers of volume being balanced against each other,
\bea
V & = & A e^{-\lambda_1 \Phi} - B e^{-\lambda_2 \Phi},
\label{potpert} \\
V^{'} & = & -\lambda_1 A e^{-\lambda_1 \Phi} + B \lambda_2 e^{- \lambda_2 \Phi},
\eea
and so at the minimum $\lambda_1 A e^{-\lambda_1 \Phi} = B \lambda_2 e^{-\lambda_2 \Phi}$ and $V \vert_{min} = \left(
\frac{\lambda_2}{\lambda_1} - 1 \right) B e^{-\lambda_2 \Phi}$.
From this we can derive
\be
V^{''}\vert_{min} = \left( \lambda_1^2 A e^{-\lambda_1 \Phi} - \lambda_2^2 B e^{-\lambda_2 \Phi} \right)\vert_{min} = -\lambda_1 \lambda_2 V \vert_{min},
\ee
and also
\be
V^{(n)} \vert_{min} = (-1)^{n-1} \lambda_1 \lambda_2  \frac{\left( \lambda_1^{n-1} - \lambda_2^{n-1} \right)}{\lambda_1 - \lambda_2} V \vert_{min}.
\label{ddee}
\ee
As for LVS, this depends only on the powers $\lambda_1$ and $\lambda_2$ and not on the microphysics such as $A$ or $B$.
In general, Eq. (\ref{potpert}) will have a minimum at small or moderate values of the volume and so is not directly comparable to LVS. In the limit where
$\lambda_1 \simeq \lambda_2$, this minimum moves out to large volumes. Writing $\lambda_2 = \lambda_1 + (\lambda_2 - \lambda_1)$ and expanding perturbatively, Eq. (\ref{ddee}) becomes
\be
V^{(n)} \vert_{min} = (-1)^{n-1} (n-1) \lambda_1^{n-1} V \vert_{min} + \mc{O}(\lambda_2 - \lambda_1),
\ee
exactly matching the LVS result. It is appealing that we obtain the same answer as for LVS despite the different starting point, but note that this should not be too surprising as in the limit of $\lambda_2 \to \lambda_1$  Eq. (\ref{potpert}) takes a very similar form to that used previously in Eq. (\ref{hij}).

\subsection{KKLT}

Let us also consider the widely-studied KKLT vacuum to examine any relevant differences. The KKLT AdS vacuum also comes from IIB flux models in which the dilaton and complex stucture moduli are stabilised supersymmetrically \cite{kklt}. The effective supergravity theory for the K\"ahler moduli is then
\bea
K & = & - 3 \ln (T + \bar{T}) \nn \\
W & = & W_0 + A e^{-a T}.
\eea
This can be easily solved to obtain for a supersymmetric minimum $D_T W = 0$. While the above represents the simplest 1-modulus model,
the general structure of KKLT can also be applied to models with many moduli, in which a supersymmetric minimum comes from balancing non-perturbative superpotential terms against the constant $W_0$. To obtain minima at moderate values of volume which one hopes to be in the supergravity limit, $W_0$ is tuned small using fluxes.

As KKLT is a supersymmetric minimum, it follows that at the minimum $V_{AdS} = - 3m_{3/2}^2 M_P^2$ and so $R_{AdS} = \frac{1}{m_{3/2}}$.
The mass of the volume modulus, however, satisfies \cite{ChoiNilles}
\be
m_{\Phi} \sim m_{3/2} \ln \left( \frac{M_P}{m_{3/2}} \right).
\ee
This implies that the conformal dimension of the dual operator in KKLT satisfies\footnote{Notice that in \cite{14126999} this logarithmic dependence was not spelled out and the conformal dimension was taken to be of $\mathcal{O}(1)$.}
\be
\Delta (\Delta - 3) \sim \left( \ln \left( \frac{M_P}{m_{3/2}} \right) \right)^2 \sim \left( \ln R_{AdS} \right)^2,
\ee
and so in the limit of small $W_0$ (or small $m_{3/2}$ in KKLT) one obtains $\Delta \sim \ln \left( \frac{M_P}{m_{3/2}} \right)$.
In KKLT control comes from tuning of $W_0$ to small values, and so in the maximally controlled limit of arbitrary fine-tuning
of $W_0 \to 0$ the conformal dimensions of the low-lying scalar operators grow as $\ln \left( M_P/ m_{3/2} \right)$ with no bound.
In contrast to LVS, KKLT therefore does not seem to offer a holographic interpretation as a
perturbation about a particular CFT obtained in the control limit (here $W_0 \ll 1$).

As stabilisation is supersymmetric, this mass scale also applies to both the axion partner of the volume modulus and the
fermionic partners within the supersymmetry multiplet. In the limit of small $W_0$, the dimensions of these low-lying operators then grow as
$\ln \left( R_{AdS}/ l_{Pl} \right)$, where $l_{Pl}$ is the 4-dimensional Planck length.

In KKLT the gravitino, with a mass of $m_{3/2}$, is the lightest massive excitation. In that decoupling of the scalar modes is possible (as any such decoupling is only logarithmic in $W_0$, and the difficulty of tuning fluxes scales as $\vert W_0 \vert^2$ \cite{DenefDouglas}, it is unclear whether these can ever be regarded as parametrically decoupled), a dual of KKLT would reduce to a generalised free field CFT consisting of the $\mc{N}=1$ multiplet of the stress-tensor (dual to $g_{\mu \nu}$) and a spin-3/2 operator dual to the gravitino.

\section{Bootland Conjectures}
\label{BootlandSec}

We now come to the more conjectural part of this paper: the idea that swampland constraints on the consistency of AdS quantum gravity theories will be equivalent, via holography, to the bootstrap consistency constraints of unitarity and crossing symmetry on the dual 3d CFT (for discussions of the weak gravity conjecture in AdS/CFT see \cite{150901647, 160309745}). Unitarity constraints have been used to 
constrain the space of consistent effective field theories (for example see \cite{0602178}) and these ideas have also been used in connection with the swampland for gravitational effective field theories \cite{14077865, 181003637}

While we propose this as a conjecture, we want to use the effective LVS AdS theory to provide motivational arguments for it.
Let us re-state the form of the AdS interactions between the two light scalar fields $\delta \Phi$ and $a$ present in the effective field theory of LVS,
\bea
{\mc L}_{(\delta \Phi)^n} & = & (-1)^{n-1} \lambda^n (n-1)  \left( - 3 \frac{M_P^2}{R_{AdS}^2} \right) \frac{1}{n!} \left( \frac{\delta \Phi}{M_P} \right)^n \left( 1 + \mc{O} \left( \frac{1}{\lambda \langle \Phi \rangle} \right) \right), \label{LVSsummary1}  \\
\mc{L}_{(\delta \Phi)^{n-2} a a } & = & \left( - \sqrt{\frac{8}{3}} \right)^{(n-2)} \frac{1}{2 (n-2)!}\left( \frac{\delta \Phi}{M_P}\right)^{n-2} \partial_{\mu} a \partial^{\mu} a,
\label{LVSsummary}
\eea
with $\lambda = \sqrt{\frac{27}{2}}$ (there are in addition the heavier modes whose dimensions diverge in the $\mc{V} \to \infty$ limit).

If one takes the LVS solution seriously, then it defines a solution to quantum gravity on AdS space with a low-energy spectrum and interactions that are, in the $\mc{V} \to \infty$ limit, radiatively exact and entirely specified in terms of $R_{AdS}$. Using the holographic logic of AdS/CFT, this AdS solution determines CFT correlators via a Witten diagram expansion in AdS (for pedagogical discussions, see \cite{11014163, 160804948}). Assuming LVS to be correct, we then regard the structure of Eq. (\ref{LVSsummary}) together with the dimensions $\Delta_{\Phi}$ and $\Delta_a$ as defining a dual CFT, $CFT_{LVS}$ (or at least the part of the dual CFT corresponding to the low-dimension operators).

Such CFTs are constrained by the consequence of unitarity and conformal symmetry.
Recent years have seen great progress in understanding the implications of these constraints, in particular
using the techniques of the conformal bootstrap. This has led to powerful results on (for example) the value of the lowest lying allowed operator dimensions and, given certain assumptions about which operators are relevant, the uniqueness of the 3d Ising model \cite{14064858}.
One can view the set of CFT properties forbidden by the conformal bootstrap as defining a CFT swampland. We therefore want to conjecture that these constraints are, under AdS/CFT, equivalent --- namely, that swampland constraints on consistent AdS theories of quantum gravity can be reinterpreted as an inability of a dual field theory to satisfy both unitarity and crossing symmetry.\footnote{While this mainly refers to the low energy spectrum, one can similarly conjecture that the presence of the tower of string and KK modes are, holographically, essential to satisfy generalised modular invariance of the 3d CFT.}

\begin{conjecture}
Modifications to Eqs. (\ref{LVSsummary1}) and  (\ref{LVSsummary}) that place the resulting AdS theory in the swampland are equivalent to modifications to the dual field theory that make it unable to satisfy the CFT bootstrap constraints of unitarity and crossing symmetry.
\end{conjecture}
We can see this as a special version of a more general conjecture,
\begin{conjecture2}
Swampland constraints on consistent AdS theories of quantum gravity are equivalent to bootstrap constraints on consistency of the
dual CFT.
\end{conjecture2}

To have teeth, conjectures need to be translatable into quantitative statements.
To illustrate this, we first show ways of modifying the interactions of Eqs. (\ref{LVSsummary1}) and (\ref{LVSsummary}) that appear
plausible in field theory but which, based on our knowledge of string theory compactified down to four dimensions, we expect to be physically nonsensical and so part of the swampland.
\begin{enumerate}
\item
In Eq. (\ref{LVSsummary}), the axion kinetic term comes from $e^{-\sqrt{\frac{8}{3}} \Phi / M_P}$ and behaves as $\frac{M_P}{\mc{V}^{2/3}}$ in the large volume limit.
A modification to Eq. (\ref{LVSsummary}) such that the axion kinetic term either remained constant
or increased in the limit $\mc{V} \to \infty$ would imply that the volume of moduli space in the $\mc{V} \to \infty$ limit would diverge.
Such behaviour is strongly believed to be incompatible with quantum gravity (e.g. \cite{0303252, OoguriVafaI}, and
also see \cite{180208264, 180208698} for recent discussions of kinetic terms in a decompactification limit from field theory perspectives).

This implies, for example, that a replacement of $e^{-\sqrt{\frac{8}{3}} \Phi}$ by $e^{+\sqrt{\frac{8}{3}} \Phi}$
in Eq. (\ref{akineticterm}) --- equivalently, an additional factor of $(-1)^n$ in Eq. (\ref{LVSsummary}) --- should be inconsistent.
More generally, we also expect that \emph{any} modification of the axion kinetic term in Eq. (\ref{LVSsummary}) to
originate from a positive-exponential couplings such as $e^{+ \alpha \Phi} \partial_{\mu} a \partial^{\mu} a$ (or even simply $\partial_{\mu} a \partial^{\mu} a$ with no $\Phi$ dependence) would be forbidden. We expect this to hold
despite the fact that the volume axion $a$ is simply a spectator field for the LVS dynamics and the $a$ kinetic term plays no explicit role in LVS.

\item
The LVS potential scales as $\mc{V}^{-3}$ in the large-volume limit, and this factor of $\mc{V}^{-3}$ corresponds to $\lambda = \sqrt{\frac{27}{2}}$ within Eq. (\ref{LVSeffPotential}). Based on what we know about perturbative quantum corrections within string theories, it would appear impossible to generate an LVS-esque potential as in Eq. (\ref{effPotential}), but with a scaling of (say)
$\mc{V}^{-100}$ replacing the $\mc{V}^{-3}$ as the prefactor that gives the dominant power in the $\mc{V} \to \infty$ limit.
In such a case, we know of no way to prevent more dominant effects at lower powers of volume arising from string loops.

This suggests that the structure of Eqs. (\ref{LVSsummary1}) and (\ref{LVSsummary}), certainly in the limit $\mc{V} \gg 1$, would be inconsistent with quantum gravity when $\lambda \gg \sqrt{\frac{27}{2}}$.

\item
A much stronger form of this is the claim that, not only is $\lambda \gg \sqrt{\frac{27}{2}}$ forbidden, but that LVS is the \emph{only} possible mechanism for producing scale-separated vacua at $\mc{V} \ggg 1$ (modulo constructions such as \cite{08041248} that are morally equivalent) --- i.e., the delicate balancing of $\alpha'$ and instanton effects which leads to an exponentially large volume minimum is only possible for an asymptotic $\mc{V}^{-3}$ scaling. This much stronger form of the previous conjecture would then state that $\lambda = \sqrt{\frac{27}{2}}$ is the \emph{only} consistent value in Eq. (\ref{LVSsummary1}).

\item
Within string theory, the presence of exponentials in the canonical field $\Phi$, that expand to give all the individual interactions, is `obvious' --- they arises from effects that are power-law in the physical volume and so are exponential in the canonical field $\Phi$.
In this respect, it appears unavoidable to have an axionic kinetic term that is exponential in $\Phi$.
However, when writing effective low-energy Lagrangians in quantum field theory there seems to be no obvious perturbative problem with (for example) changing by a factor of two the coefficient of the (say) $\left( \delta \Phi / M_P \right)^{100} \partial_{\mu} a \partial^{\mu} a$ term in the Lagrangian --- or indeed any number of other such discrete changes in the coefficients of terms
in Eq. (\ref{LVSsummary1}) and Eq. (\ref{LVSsummary}).

Based on the string theory origin, though, we claim that any such number of discrete changes in the
interactions in Eqs. \ref{LVSsummary1} and \ref{LVSsummary} would be inconsistent with quantum gravity.
\end{enumerate}

We now want to argue that the requirements of crossing symmetry and unitarity of $CFT_{LVS}$ will be sensitive to such structures, although
attempts at a detailed analysis of the Witten diagrams or construction of a dual CFT are beyond the scope of this paper.

To be self-contained, we first briefly review the requirement of crossing symmetry within a CFT. In a CFT, operators can be organised in times of primaries and their descendants. As any two primary operators $\mc{O}_a$ and $\mc{O}_b$ approach each other, they can be replaced by an operator product expansion
\be
\mc{O}_a(x_2) \mc{O}_b(x_1) = \sum_i C_{ab}^i \vert x_2 - x_1 \vert^{\Delta_i - \Delta_a - \Delta_b} \mc{O}_i(x_1) + {\rm descendants},
\ee
where $\Delta_a$ is the conformal dimension of operator $\mc{O}_a$, the index $i$ sums over all operators of the CFT and the descendant contributions are related by kinematics to that of the primary operators. Given the dimensions of the
primary operators, the structure constants $C_{ab}^i$ (which must be real in a unitary theory) then completely define the CFT, and in principle any higher point CFT correlator can be evaluated through a recursive series of OPE expansions down to sums over lower-point correlators.

The result of this recursive evaluation must be independent of the order in which the OPEs are performed.
In particular, this holds for a CFT 4-point function $\langle \mc{O}_a(x_1) \mc{O}_b(x_2) \mc{O}_c(x_3) \mc{O}_d(x_4) \rangle$.
There are two distinct ways to expand the 4-point function --- either by first performing the
$\mc{O}_a(x_1) \mc{O}_b(x_2)$ and $\mc{O}_c(x_3) \mc{O}_d(x_4)$ OPEs and then contracting the resulting 2-pt functions in the `s-channel', or equivalently by first expanding $\mc{O}_a(x_1) \mc{O}_c(x_3)$ and $\mc{O}_b(x_2) \mc{O}_d(x_4)$ OPEs and contracting in the `t-channel'.
As in principle all fields can appear within the OPE, this gives the schematic expression
\be
\sum_i C_{ab}^{i} \, C_{cd}^{i} \, \mc{G}_{i}^{12 \to 34}(x_1, x_2, x_3, x_4) = \sum_j C_{ac}^{j}
\, C_{bd}^{j} \, \mc{G}_{j}^{13 \to 24} (x_1, x_3, x_2, x_4).
\ee
Here $i$ and $j$ sum over all fields in the CFT, $C_{ab}^{i}$ represents the structure constant for $\mc{O}_i$ within the OPE of $\mc{O}_a$ and $\mc{O}_b$, and $\mc{G}_{i}^{12 \to 34}$ represents the effects of exchange of field $\mc{O}_i$ (plus descendants) from $x_1$ to $x_3$, arising from the OPEs of $\mc{O}_a(x_1) \mc{O}_b(x_2)$ and $\mc{O}_c(x_3) \mc{O}_d(x_4)$.

Now let us specialise to CFTs that would be dual to any LVS (or related) construction. As described in section \ref{LVS2pt}, the low-lying single trace operators will consist of the stress tensor plus an even parity scalar field dual to the volume modulus (which we also denote $\Phi$) and an odd-parity scalar field dual to the volume axion (which we also denote $a$). Such $\mbb{Z}_2$ symmetric 3d CFTs have been analysed extensively within the bootstrap literature, but mainly from a perspective where both scalar operators would be relevant (e.g. see \cite{14064858, 170805718}).

We assume that the low-lying sector should satisfy crossing symmetry on its own accord (i.e. without having to include the effects of heavy states) and want to consider the structure of crossing symmetry for the $\langle \Phi(x_1) \Phi(x_2) a(x_3) a(x_4) \rangle$ correlator,
\be
\sum_i C_{\Phi \Phi}^{i} \, C_{a a}^{i} \, \mc{G}_{i}^{12 \to 34}(x_1, x_2, x_3, x_4) = \sum_j (C_{\Phi a}^{j})^2 \,
 \, \mc{G}_{j}^{13 \to 24}(x_1, x_3, x_2, x_4).
\label{LVScrossing}
\ee
In the asymptotic large volume limit, the conformal dimensions of the operators are fixed --- they arise from the single trace fields $a$, $\Phi$ and $g_{\mu \nu}$, plus double- and higher-trace modes such as $\Phi^2$, $\Phi a$, etc. The interesting physics is therefore in the structure
constants $C_{ab}^{i}$, which can in principle be found through Witten diagrams evaluated using the interactions of the AdS theory.

The $\mbb{Z}_2$ parity allows the operators to be divided into even- and odd-parity sets, $\mc{O}^{i+}$ (which includes the identity operator
${\bf 1}$) and $\mc{O}^{j-}$. In Eq. (\ref{LVScrossing}) the left-hand side involves exchange of $\mc{O}^{i+}$ and the right-hand side the exchange of $\mc{O}^{j-}$.
The OPE expansions take the schematic form
\bea
\mc{O}^{i+} \ti \mc{O}^{i+} & \sim & C_{i+ i+}^{i+} \mc{O}^{i+} , \nn \\
\mc{O}^{i+} \ti \mc{O}^{j-} & \sim & C_{i+ j-}^{j-} \mc{O}^{j-} , \nn \\
\mc{O}^{j-} \ti \mc{O}^{j-} & \sim & C_{j- j-}^{i+} \mc{O}^{i+} ,
\label{OPEexpansions}
\eea
and so the LHS of Eq. (\ref{LVScrossing}) involves the first and third of these OPEs while the RHS only involves the second.
Equality of the left and right hand sides of Eq. (\ref{LVScrossing})
will then imply relationships between these different terms.

From Eqs. (\ref{LVSsummary1}) and (\ref{LVSsummary}), there are two basic forms of scalar field interactions --- the
potential terms which depend solely on the even-parity $\Phi$ field and include $\Phi^3, \Phi^4, \ldots$ interactions, and the kinetic terms involving mixed even-odd interactions of the form $\Phi^n \partial_{\mu} a \partial^{\mu} a$.
We expect the $\mbb{Z}_2$ symmetry to restrict the origin of the various OPE coefficients: for example, we expect the potential terms
 to be responsible for $C_{\Phi \Phi}^{\Phi^m}$ terms, whereas the $C_{a a}^{i+}$ coefficients will arise from the axion kinetic terms.

For example, on the right hand side of Eq. (\ref{LVScrossing}) the OPEs involve $\left( C_{\Phi a}^{a} \right)^2$,
which in a unitary theory must be positive as $C_{\Phi a}^{a}$ is real. However, on the left hand side the OPEs involve the product $C_{\Phi \Phi}^{\Phi} C_{a a}^{\Phi}$. Here the sign is not fixed \emph{a priori} and the contribution of this term will be sensitive to the signs and coefficients of both the $\Phi^3$ self-interaction and also
the $\Phi \partial_{\mu} a \partial^{\mu} a$ axion kinetic term.\footnote{This is one reason why we explicitly include unitarity as part of our conjectures; a Witten diagram might respect crossing symmetry but not unitarity (e.g. see \cite{11014163}).}
While the full statement of crossing symmetry requires the sum over all intermediate modes in Eq. (\ref{LVScrossing}), this suggests the presence of intricate consistency requirements on the interaction coefficients of Eq. (\ref{OPEexpansions}).
Assuming that Eq. (\ref{LVSsummary1}) and Eq. (\ref{LVSsummary}) do define a theory $CFT_{LVS}$, the essence of our conjectures is that
modifications of Eq. (\ref{LVSsummary1}) and Eq. (\ref{LVSsummary}) into a swampland theory will correspond to modifying
the OPE structure constants of $CFT_{LVS}$ such that it can no longer represent a unitary conformal field theory.

Let us be a little more specific on this and consider one specific modifications of the axion kinetic term
while leaving the LVS scalar potential unaffected. In Eqs (\ref{LVSsummary1}) and (\ref{LVSsummary}), the coefficient of the $\Phi^3$ and $\Phi \partial_{\mu} a \partial^{\mu} a$ terms are both negative, and so the product has positive sign (cf our discussion of $C_{\Phi \Phi}^{\Phi} C_{a a}^{\Phi}$).
We now imagine changing the sign of the exponential in the $e^{-\sqrt{\frac{8}{3}} \Phi} \partial_{\mu} a \partial^{\mu} a$ kinetic term to
$e^{+\sqrt{\frac{8}{3}} \Phi} \partial_{\mu} a \partial^{\mu} a$, modifying Eq. (\ref{LVSsummary}) by a factor of $(-1)^n$.
Based on everything that is known about string theory, this new theory is in the swampland.
We expect this modification to change $C_{a a}^{i+}$ couplings by similar factors of $(-1)^n$, which would significantly modify
the signs of the OPE sum on the LHS of Eq. (\ref{LVScrossing}) while leaving the RHS unaffected.
More generally, one can imagine continuously varying the coefficient in the exponential from $\sqrt{\frac{8}{3}}$ to $-\sqrt{\frac{8}{3}}$, which would be a trajectory in the AdS theory that leads us from the landscape into the swampland. By continuity there is a definite point on this trajectory where the AdS theory enters the swampland, and so we expect that some similar marker should occur on the CFT side at this point.

We stress that these arguments do not represent
full calculations of the CFT structure constants starting from the AdS theory.
In this paper we are simply motivating why CFT structures should be sensitive to modifications of consistent non-swampland
AdS theories into inconsistent swampland AdS theories --- it would be striking if the swampland boundary on the AdS side was not mirrored
by a similar boundary on the CFT side.

\section{Conclusions}

Holographic descriptions of flux compactifications and moduli stabilisation has been a subject of intermittent interest over the years. We believe now is a suitable time to revisit this topic in detail, following the technological advances in the study of the
conformal bootstrap.
We hope to have shown that holographic duals of flux compactifications may not be as far-fetched as has been thought, as the effective low-energy Lagrangian of the Large Volume Scenario is, when rewritten in terms of $R_{AdS}$, uniquely determined in the large volume limit.
This rewritten form is remarkably simple, radiatively stable and universal in the limit of $\mc{V} \to \infty$.
In particular, there is a sharp prediction of $\Delta_{\Phi} = \frac{3}{2} \left( 1 + \sqrt{19} \right)$ for the dimension of the operator dual to the volume modulus. This also offers the chance to disprove the Large Volume Scenario by showing, from the CFT side, an inconsistency in the structure of Eqs. (\ref{LVSsummary1}) and (\ref{LVSsummary}).

There are various modifications that can be made to the low-energy Lagrangian of Eqs. (\ref{LVSsummary1}) and (\ref{LVSsummary}) that, from an AdS perspective, would put the theory rather clearly into the swampland.
We have conjectured that these modifications should have an interpretation on the CFT side in terms of a modification to the field theory that
makes it impossible to satisfy the conformal field theory constraints of unitarity and crossing symmetry. We have also conjectured that such bootland constraints will also hold more generally for other AdS moduli stabilisation scenarios.

Of course, it may be that such a relationship is false, and inconsistent modifications to AdS Lagrangians still lead to (at least apparently)
consistent CFTs. In this case it would not be possible to draw any conclusions about the swampland through CFT arguments. On the other hand, if
it were possible to identify CFT inconsistencies for certain couplings in the effective AdS Lagrangian, then this would open the door to potential disproofs of standard moduli stabilisation scenarios such as LVS - or at least to far more robust checks on their consistency. The ideal result would be a clear map of consistent and inconsistent AdS Lagrangians, within which one could locate the standard moduli stabilisation Lagrangians.%%JC

More generally, the simplicity and universality of the rewritten LVS Lagrangian
may also offer a new approach to thinking about swampland constraints on consistent AdS gravity theories, potentially entering a new era of `quantum gravitational string phenomenology' \cite{181007673}.

\section*{Acknowledgments}

We thank Jo\~{a}o Penedones and David Simmons-Duffin for clarifications, and David Andriot, Shanta de Alwis, Anamar{\'{i}}a Font, Rajesh Gupta, Francesco Muia, Fabian Ruehle, Julian Sonner, Cumrun Vafa and Ben Withers for discussions. We also thank Nick Dorey for suggesting the expression `bootland'. JC is supported by the STFC Oxford Consolidated Grant.

\bibliography{lvscft}{}
\bibliographystyle{JHEP}
\end{document}